\let\useblackboard=\iftrue
%
%
\newfam\black
\input harvmac.tex
\noblackbox
\def\Title#1#2{\rightline{#1}
\ifx\answ\bigans\nopagenumbers\pageno0\vskip1in%
\baselineskip 15pt plus 1pt minus 1pt
\else
\def\listrefs{\footatend\vskip 1in\immediate\closeout\rfile\writestoppt
\baselineskip=14pt\centerline{{\bf References}}\bigskip{\frenchspacing%
\parindent=20pt\escapechar=` \input
refs.tmp\vfill\eject}\nonfrenchspacing}
\pageno1\vskip.8in\fi \centerline{\titlefont #2}\vskip .5in}

\ifx\answ\bigans\def\tcbreak#1{}\else\def\tcbreak#1{\cr&{#1}}\fi
\useblackboard
\message{If you do not have msbm (blackboard bold) fonts,}
\message{change the option at the top of the tex file.}

\font\blackboard=msbm10 scaled \magstep1
\font\blackboards=msbm7
\font\blackboardss=msbm5
\textfont\black=\blackboard
\scriptfont\black=\blackboards
\scriptscriptfont\black=\blackboardss

\else

\fi
%
\def\yboxit#1#2{\vbox{\hrule height #1 \hbox{\vrule width #1
\vbox{#2}\vrule width #1 }\hrule height #1 }}
\def\fillbox#1{\hbox to #1{\vbox to #1{\vfil}\hfil}}
\def\ybox{{\lower 1.3pt \yboxit{0.4pt}{\fillbox{8pt}}\hskip-0.2pt}}

\def\comments#1{}

\def\eps{\epsilon}
\def\half{{1\over 2}}
\def\Tr{{{\rm Tr\  }}}

\def\CN{{\cal N}}

\def\a{\alpha}

\def\II{\relax{I\kern-.07em I}}
\def\IIA{{\II}A}
\def\IIB{{\II}B}

\def\inbar{\,\vrule height1.5ex width.4pt depth0pt}
\def\IZ{\relax\ifmmode\mathchoice
{\hbox{\cmss Z\kern-.4em Z}}{\hbox{\cmss Z\kern-.4em Z}}
{\lower.9pt\hbox{\cmsss Z\kern-.4em Z}}
{\lower1.2pt\hbox{\cmsss Z\kern-.4em Z}}\else{\cmss Z\kern-.4em
Z}\fi}
\def\IB{\relax{\rm I\kern-.18em B}}
\def\IC{{\relax\hbox{$\inbar\kern-.3em{\rm C}$}}}
\def\ID{\relax{\rm I\kern-.18em D}}
\def\IE{\relax{\rm I\kern-.18em E}}
\def\IF{\relax{\rm I\kern-.18em F}}
\def\IG{\relax\hbox{$\inbar\kern-.3em{\rm G}$}}
\def\IGa{\relax\hbox{${\rm I}\kern-.18em\Gamma$}}
\def\IH{\relax{\rm I\kern-.18em H}}
\def\IK{\relax{\rm I\kern-.18em K}}
\def\IP{\relax{\rm I\kern-.18em P}}
\def\pp{{\relax{=\kern-.42em |\kern+.2em}}}

\font\cmss=cmss10 \font\cmsss=cmss10 at 7pt
\def\IR{\relax{\rm I\kern-.18em R}}

\def\Tr{{\rm Tr\ }}


%
%

\def\NP{{\it Nucl. Phys.\ }}
\def\AP{{\it Ann. Phys.\ }}
\def\PL{{\it Phys. Lett.\ }}
\def\PR{{\it Phys. Rev.\ }}
\def\PRL{{\it Phys. Rev. Lett.\ }}

\def\JMP{{\it J. Math. Phys.\ }}

\def\IJMP{{\it Int. Jour. Mod. Phys.\ }}
\def\Mod{{\it Mod. Phys. Lett.\ }}

\def\JGP{{\it J. Geom. Phys.\ }}
\def\ZP{ {\it Z. Phys.\ }}

\Title{ \vbox{\baselineskip12pt\hbox{hep-th/9709015}
\hbox{BROWN-HET-1091}
}}
{\vbox{
\centerline{$E_8 \times E_8$ Small Instantons in Matrix Theory}}}

\centerline{David A. Lowe}
\medskip
\centerline{Department of Physics}
\centerline{Brown University}
\centerline{Providence, RI 02912, USA}
\centerline{\tt lowe@het.brown.edu}
\bigskip
A formulation of new six-dimensional theories with $(1,0)$
supersymmetry and $E_8$ global symmetry is proposed. The
model is based on the large $n$ theory describing $n$ D-strings
interacting
with parallel D-fivebranes in Type I string theory.

\Date{August, 1997}

\lref\sen{A. Sen, ``A Note on Marginally Stable Bound States in Type
II String Theory,'' \PR {\bf D54} (1996) 2964; hep-th/9510229.}
\lref\daniel{U. Danielsson and G. Ferretti, ``The Heterotic Life of
the D-particle,'' \IJMP {\bf A12} (1997) 4581, hep-th/9610082.}
\lref\polwit{J. Polchinski and E. Witten, ``Evidence for
Heterotic-Type I String Duality,'' \NP {\bf B460} (1996) 525,
hep-th/9510169.}
\lref\kacsil{S. Kachru and E. Silverstein, ``On Gauge Bosons in the
Matrix Model Approach to M Theory,'' \PL {\bf B396} (1997) 70, hep-th/9612162.}
\lref\danieltwo{U. Danielsson, G. Ferretti and B. Sundborg,
``D-particle Dynamics and Bound States,'' \IJMP {\bf
A11} (1996) 5463, hep-th/9603081.}
\lref\bfss{T. Banks, W. Fischler, S. Shenker and L. Susskind,
``M-Theory as a Matrix Model: A Conjecture,'' hep-th/9610043.}
\lref\horava{P. Horava and E. Witten, ``Heterotic and Type I String
Dynamics from Eleven Dimensions,'' \NP {\bf B460} (1996) 506,
hep-th/9510209.}
\lref\dai{P. Horava, ``Strings on Worldsheet Orbifolds,'' \NP {\bf B327}
(1989) 461; J. Dai, R.G. Leigh and J. Polchinski, ``New Connections
Between String Theories,'' \Mod {\bf A4} (1989)
2073;  P. Horava, ``Background Duality of Open-String Models,''
\PL {\bf B231} (1989) 251.}
\lref\gresch{M. Green, J. Schwarz and E. Witten, ``Superstring
Theory,''
Cambridge University Press, 1987.}
\lref\oldzero{M. Claudson and M.B. Halpern, ``Supersymmetric Ground
State Wave Functions,'' \NP {\bf B250} (1985) 689;
M. Baake, P. Reinicke and V. Rittenberg, ``Fierz Identities for Real
Clifford Algebras and the Number of Supercharges,'' \JMP {\bf 26} (1985) 1070;
R. Flume, ``On Quantum Mechanics with Extended Supersymmetry and
Nonabelian Gauge Constraints,'' \AP {\bf 164} (1985) 189.}
\lref\doug{M.R. Douglas, D. Kabat, P. Pouliot and S.H. Shenker,
``D-branes and Short Distances in String Theory,'' \NP {\bf B485}
(1997) 85, hep-th/9608024.}
\lref\kabat{D. Kabat and P. Pouliot, ``A Comment on Zerobrane Quantum
Mechanics,'' \PRL {\bf 77} (1996) 1004,
hep-th/9603127.}
\lref\hoppe{B. de Wit, J. Hoppe, H. Nicolai, ``On the Quantum
Mechanics of Supermembranes,'' \NP {\bf B305} (1988)
545.}
\lref\lowe{D.A. Lowe, ``Bound States of Type I$'$ D-particles and
Enhanced Gauge Symmetry,'' \NP {\bf B501} (1997) 134, hep-th/9702006.}
\lref\dbranes{J. Polchinski, ``TASI Lectures on D-Branes,''
hep-th/9611050; J. Polchinski, S. Chaudhuri and C. Johnson,
``Notes on D-Branes,'' hep-th/9602052.}
\lref\duality{E. Witten, ``String Theory Dynamics in Various
Dimensions,'' \NP {\bf B443} (1995) 85, hep-th/9503124;
C. Hull and P. Townsend ``Unity of Superstring Dualities,'' \NP {\bf
B438} (1995) 109, hep-th/9410167.}
\lref\mtheory{J. Schwarz, ``The Power of M-Theory,'' \PL {\bf B367}
(1996) 97, hep-th/9510086.}
\lref\green{M. Green and J. Schwarz, ``The hexagon gauge anomaly in
Type I superstring theory,'' \NP {\bf B255} (1985) 93.}
\lref\bss{T. Banks, S. Seiberg and E. Silverstein, ``Zero and
One-dimensional Probes with N=8 Supersymmetry,'' hep-th/9703052.}
\lref\moore{R. Dijkgraaf, G. Moore, E. Verlinde and H. Verlinde,
``Elliptic Genera of Symmetric Products and Second Quantized
Strings,'' hep-th/9608096.}
\lref\dvv{R. Dijkgraaf, E. Verlinde and H. Verlinde, ``Matrix String
Theory,'' hep-th/9703030.}
\lref\banksei{T. Banks and N. Seiberg, ``Strings from Matrices,''
hep-th/9702187.}
\lref\motl{L. Motl and L. Susskind, ``Finite N Heterotic Matrix Models
and Discrete Light Cone Quantization,'' hep-th/9708083.}
\lref\witbound{E. Witten, ``Bound States of Strings and p-Branes,''
\NP {\bf B460} (1996) 335,
hep-th/9510135.}
\lref\papadop{G. Papadopoulos and P.K. Townsend,  ``Kaluza-Klein on
the Brane,'' \PL {\bf B393} (1997) 59, hep-th/9609095.}
\lref\banks{T. Banks and L. Motl, ``Heterotic Strings from Matrices,''
{\it JHEP} {\bf 12} (1997) 004, hep-th/9703218.}
\lref\srey{N. Kim and S.J. Rey, ``M(atrix) Theory on an Orbifold and
Twisted Membrane,'' hep-th/9701139.}
\lref\dougone{M.R. Douglas, ``Branes within Branes,''
hep-th/9512077.}
\lref\itoyama{Y. Arakane, H. Itoyama, H. Kunitomo and A. Tokura,
``Infinity Cancellation, Type I$'$ Compactification and
String S-Matrix Functional,'' hep-th/9609151.}
\lref\witsmall{E. Witten, ``Small Instantons in String Theory,'' \NP
{\bf B460} (1996) 541, hep-th/9511030.}
\lref\adhm{M.F. Atiyah, V.G. Drinfeld, N.J. Hitchin and Y.I. Manin,
``Construction of Instantons,'' \PL {\bf A65} (1978) 185.}
\lref\corgod{E. Corrigan and P. Goddard, ``Construction of Instanton
and
Monopole solutions and Reciprocity,'' \AP {\bf 154} (1984) 253.}
\lref\christ{N.H. Christ, E.J. Weinberg and N.K. Stanton, ``General
Self-Dual Yang-Mills Solutions,'' \PR {\bf D18} (1978) 2013.}
\lref\ganorhan{O.J. Ganor and A. Hanany, ``Small $E_8$ Instantons and
Tensionless Noncritical Strings,'' \NP {\bf B474} (1996) 122,
hep-th/9602120.}
\lref\vafa{C. Vafa, ``Evidence for F Theory,'' \NP {\bf B469} (1996)
403, hep-th/9602022.}
\lref\vafamor{D.R. Morrison and C. Vafa, ``Compactifications of F Theory
on Calabi-Yau Threefolds I and II,'' \NP {\bf B473} (1996) 74,
hep-th/9602114; \NP {\bf B476} (1996) 437, hep-th/9603161.}
\lref\wittensei{N. Seiberg and E. Witten, ``Comments on String
Dynamics
in Six Dimensions,'' \NP {\bf B471} (1996) 121, hep-th/9603003.}
\lref\witphase{E. Witten, ``Phase Transitions in M-Theory and
F-Theory,''
\NP {\bf B471} (1996) 195, hep-th/9603150.}
\lref\sethi{S. Sethi and M. Stern, ``A comment on the spectrum of
H-monopoles,'' \PL {\bf B398} (1997) 47, hep-th/9607145.}
\lref\witstrong{E. Witten, ``Physical Interpretation of Certain Strong
Coupling Singularities,'' \Mod {\bf A11} (1996) 2649, hep-th/9609159.}
\lref\ganor{O.J. Ganor, ``A Test of the Chiral $E_8$ Current Algebra
on a 6d Noncritical String,'' \NP {\bf B479} (1996) 197,
hep-th/9607020.}
\lref\klemm{A. Klemm, P. Mayr and C. Vafa, ``BPS States of Exceptional
Non-Critical Strings,'' hep-th/9607139.}
\lref\dougvafa{M.R. Douglas, S. Katz and C. Vafa, ``Small Instantons,
del Pezzo Surfaces and Type I$'$ theory,'' \NP {\bf 497} (1997) 155,
hep-th/9609071.}
\lref\morseib{D.R. Morrison and N. Seiberg, ``Extremal Transitions and
Five-Dimensional Supersymmetric Gauge Theories,'' \NP {\bf B483}
(1996) 229, hep-th/9609070.}
\lref\dougtwo{M.R. Douglas, ``Gauge Fields and D-branes,''
hep-th/9604198.}
\lref\ganmo{O.J. Ganor, D.R. Morrison and N. Seiberg, ``Branes,
Calabi-Yau Spaces, and Toroidal Compactification of N=1
Six-Dimensional $E_8$ Theory,'' \NP {\bf B487} (1997) 93,
hep-th/9610251.}
\lref\intril{K. Intriligator, D.R. Morrison and N. Seiberg,
``Five-Dimensional Supersymmetric Gauge Theories and Degenerations of
Calabi-Yau Spaces,'' hep-th/9702198.}
\lref\seiberg{N. Seiberg, ``Five-Dimensional SUSY Field Theories,
Non-trivial Fixed Points and String Dynamics,'' \PL {\bf B388} (1996)
753, hep-th/9608111.}
\lref\seibergtwo{N. Seiberg, ``Non-trivial Fixed Points of the
Renormalization Group in Six Dimensions,'' \PL {\bf B390} (1997) 169,
hep-th/9609161.}
\lref\duffl{M.J. Duff, H. Lu and C.N. Pope, ``Heterotic Phase
Transitions and Singularities of the Gauge Dyonic String,'' \PL {\bf
B378} (1996) 101, hep-th/9603037.}
\lref\rey{S.J. Rey, ``Heterotic M(atrix) Strings and Their Interactions,''
\NP {\bf B502} (1997) 170, hep-th/9704158.}
\lref\lowetwo{D.A. Lowe, ``Heterotic Matrix String Theory,''
\PL {\bf B403} (1997) 243, hep-th/9704041.}
\lref\strom{A. Strominger, ``Heterotic Solitons,'' \NP {\bf B343}
(1990) 167.}
\lref\calharstr{C.G.Callan, J.A. Harvey and A. Strominger,
``World-Sheet Approach to Heterotic Instantons and Solitons,'' \NP {bf
B359} (1991) 611.}
\lref\witcom{E. Witten, ``Some Comments on String Dynamics,'' hep-th/9507121.}
\lref\chstwo{C.G.Callan, J.A. Harvey and A. Strominger, ``Worldbrane
Actions for String Solitons,'' \NP {\bf B367} (1991) 60.}
\lref\susskind{L. Susskind, ``Another Conjecture about M(atrix)
Theory,'' hep-th/9704080.}
\lref\lambert{N.D. Lambert, ``Quantizing the (0,4) Supersymmetric ADHM
Sigma Model,'' \NP {\bf B460} (1996) 221, hep-th/9508039.}
\lref\lamberttwo{N.D. Lambert, ``Heterotic p-branes from Massive Sigma
Models,''
hep-th/9605010.}
\lref\horavatwo{P. Horava, ``Matrix Theory and Heterotic Strings on
Tori,'' \NP {\bf B505} (1997) 84,
hep-th/9705055.}
\lref\berkooz{M. Berkooz and M. Rozali, ``String Dualities from Matrix
Theory,'' hep-th/9705175.}
\lref\seibergthree{N. Seiberg, ``New Theories in Six Dimensions and
Matrix Description of M-theory on $T^5$ and $T^5/\IZ_2$,'' 
\PL {\bf B408} (1997) 98,
hep-th/9705221.}
\lref\seibergsusy{N. Seiberg, ``Notes on Theories with 16
Supercharges,'' hep-th/9705117.}
\lref\dvvtwo{R. Dijkgraaf, E. Verlinde and H. Verlinde, ``5D Black
Holes and Matrix Strings,'' \NP {\bf B506} (1997) 121, hep-th/9704018.}
\lref\witadhm{E. Witten, ``Sigma Models and the ADHM Construction of
Instantons,'' \JGP {\bf 15} (1995) 215, hep-th/9410052.}
\lref\jack{R.W. Allen, I. Jack, D.R.T. Jones, ``Chiral sigma models
and the dilaton $\beta$ function,'' \ZP {\bf C41} (1988) 323.}
\lref\withig{E. Witten, ``On the conformal field theory of the Higgs
branch,'' {\it JHEP} {\bf 07} (1997) 003, hep-th/9707093.}
\lref\aharon{O. Aharony, M. Berkooz, S. Kachru, N. Seiberg and
E. Silverstein, ``Matrix Description of Interacting Theories in Six
Dimensions,'' {\it Adv. Theor. Math. Phys.} {\bf 1} (1998) 148, 
hep-th/9707079.}
\lref\aharony{O. Aharony, M. Berkooz, S. Kachru and E. Silverstein,
hep-th/9709118.}

\newsec{Introduction}

In the past few years nonperturbative effects in string theory have
been
found that are relevant for arbitrarily small string coupling
constant.
One class of such effects is the small instanton of the $SO(32)$
heterotic string theory compactified on $K3$, first studied by Witten
\witsmall.
He argued that quantum effects lead to an additional $Sp(2)$ gauge
symmetry
supported at the core of the instanton, together with a set of
massless
hypermultiplets. The
moduli space of the Higgs branch of these
new degrees of freedom matches precisely with the $SO(32)$ instanton moduli
space
obtained via
the ADHM construction  \refs{\adhm \corgod{--} \christ}.

The $E_8 \times E_8$ small instantons in heterotic string theory
on $K3$ have remained somewhat of a mystery. If we consider an
instanton in a single
$E_8$ factor,
as the instanton shrinks to
zero size there is a phase transition from a Higgs branch with 29
massless
hypermultiplets, representing the moduli of a finite size instanton
of $E_8$, to a Coulomb branch with a single massless tensor multiplet.
It has been argued the critical point is described by a tensionless
string theory in six dimensions \refs{\ganorhan, \wittensei, \duffl}.
There is evidence these theories are local quantum field theories at
non-trivial renormalization group fixed points \refs{\witstrong,
\wittensei, \seibergtwo}.

Further insight may be gained by examining these tensionless strings
from the M-theory perspective.
M theory compactified on $S^1/\IZ_2$ \horava\ has been conjectured to give
a strong-coupling description of the $E_8 \times E_8$ heterotic
string. From this point of view, the
tensionless strings discussed above arise from open membranes
stretching
between a fivebrane and one of the ``end of the world'' ninebranes of
the
$S^1/\IZ_2$ compactification \ganorhan.

Using the duality between Type
IA and M-theory on $S^1 \times S^1/\IZ_2$, we can represent
the small $E_8$ instanton by a D-fourbrane of Type IA approaching an
orientifold
plane. The full $E_8$ symmetry will be recovered in the infinite-coupling
limit when 8 D-eightbranes also
lie on the orientifold plane \seiberg. The worldvolume theory on the
fourbrane will then become one of the five-dimensional
supersymmetric non-trivial
interacting fixed-point field theories studied by Seiberg
\seiberg. \foot{The duality between the
$E_8 \times E_8$ heterotic string on $K3$ and
F theory \vafa\ on an elliptic Calabi-Yau threefold,
offers yet another means of studying the tensionless
string theories that describe small $E_8$ instantons. The small instanton
limit corresponds to the collapse of a del Pezzo surface in the
threefold
\refs{\vafamor \wittensei \ganor \klemm \dougvafa {--} \morseib}.}

A quantum formulation of M-theory on $S^1/\IZ_2$ has been recently
proposed in terms of the matrix mechanics of a system of Type IA
D-particles \refs{\daniel \kacsil {--}\lowe}. The perturbative
heterotic string may be directly recovered from this formulation
\refs{\banks, \lowetwo} and the leading order string interactions are
reproduced \lowetwo. Related results have also been obtained in \rey.

In this paper we use this heterotic matrix theory to propose for the
first time a nonperturbative
formulation of these tensionless string theories. In Type IB language,
we can realize these theories as a collection of D-strings, D-fivebranes, and
D-ninebranes, where we take $n$, the number of D-strings to infinity. This
configuration is of course T-dual to the Type IA configuration
discussed above. By keeping the radius of the $S^1$ that the D-strings
wrap finite, we may also use this model to formulate new
six-dimensional non-critical string theories which flow to the
tensionless string theories in the limit that the radius
vanishes. Such non-critical string theories have previously been
considered in the Type \II\ context in \seibergthree.
In section 2 we set up the Lagrangian describing
this configuration of D-branes. In section 3 we show how the
resulting large-$n$ gauged linear sigma model is obtained
from the original heterotic matrix model. For finite $n$ these results
may be interpreted in terms of a discretized light-cone quantization
of M-theory with the longitudinal direction
compactified on a circle of finite radius with a Wilson line.
Section 4 is devoted to a study of the spectrum of BPS states
of this theory. The Calabi-Yau approach  \refs{\ganor,\klemm}
predicts a tower of
states with $E_8$ quantum numbers. When the appropriate Wilson line is
introduced, these decompose into representations of $SO(16)$. These
results are shown to be consistent with the Matrix description of the theory.

While this work was in an advanced stage, related work proposing a
large $n$ formulation of six-dimensional theories with $(2,0)$
supersymmetry appeared \refs{\withig, \aharon}. In the context of
theories with $(1,0)$ supersymmetry, related ideas have also been
considered in \aharony.

\newsec{D-strings and D-fivebranes in Type IB}

The Lagrangian on the worldsheet of a single D-string probe in a
background of $k$ D-fivebranes and $32$ D-ninebranes was found by
Douglas \refs{\dougone, \dougtwo} to be equivalent to a linear sigma
model studied earlier by Witten \witadhm\ in the context of the ADHM
construction.
Here we will need the generalization of this to $n$ D-strings, which
introduces an additional $O(n)$ gauge symmetry.
We will follow the notation of
\dougtwo, adding $O(n)$ indices as needed.

The Lorentz group decomposes into $SO(1,1)\times SO(4)_I \times
SO(4)_E$, where $SO(4)_I$ corresponds to rotations within the
fivebranes,
while $SO(4)_E$ corresponds to rotations in the directions transverse
to
the fivebranes.
Each $SO(4)$ decomposes in turn into a product of two $SU(2)$ factors.
Doublets in the two $SU(2)$'s of $SO(4)_E$ are labeled by indices
$A$ and $Y$, while those of $SO(4)_I$ are labeled by $A'$ and $\tilde
A'$.

The fields on the worldsheet of the D-strings transform under a
$Sp(2k) \times SO(32)$ global symmetry group. The index $m$ will label the
fundamental of $Sp(2k)$ and $M$ will label the fundamental of
$SO(32)$.

The worldsheet fields are as follows
\eqn\fields{
\matrix{ {\rm boson} & {\rm fermion} & O(n) ~{\rm rep} \cr
A_\mu& \psi_+^{AA'},~\psi_+^{\tilde A' Y} & {\rm adjoint} \cr
b^{AY} & \psi_- ^{A'Y} & {\rm symmetric} \cr
b^{A' \tilde A'} & \psi_-^{A \tilde A'} & {\rm symmetric} \cr
\phi^{A' m} & \chi_-^{A m} & {\rm fundamental} \cr
 & \chi_+^{Y m} & {\rm fundamental} \cr
 & \lambda_+^M & {\rm fundamental}. \cr}
}
Here $A_\mu$ is an $O(n)$ gauge field. The $\lambda_+^M$ are real.
All the scalars and their superpartners satisfy a reality condition of
the form
\eqn\reality{
b^{AY} = \epsilon^{AB} \epsilon^{YZ} \bar b_{BZ}~.
}
$\epsilon$ is replaced by the $Sp(2k)$ invariant antisymmetric tensor
when appropriate. All indices are raised and lowered as $v_A =
\epsilon_{AB} v^B$.  The worldsheet coordinates are
denoted by $\sigma$ and $\tau$, with worldsheet metric
$ds^2=d\tau^2-d\sigma^2$, and we define light-cone coordinates
$\sigma^\pm = (\tau \pm \sigma)/\sqrt{2}$.

The configuration of D-strings and D-fivebranes is invariant under
$(0,4)$ supersymmetry, i.e. there are four real right-moving
supercharges. These supercharges satisfy a reality condition of the
form
\eqn\susreal{
Q^{AA'} = \epsilon^{AB} \epsilon^{A'B'} Q^{\dag}_{BB'}~.
}
The R-symmetry group is $SO(4)= SU(2) \times
SU(2)$. We will refer to the two different $SU(2)$ factors as $F$ and $F'$.
In the infrared the model will flow to a theory with $\CN=4$
superconformal invariance which is only invariant under a single
$SU(2)$. We will construct the model so that invariance under
$F'$ is manifest. The supersymmetry algebra is taken to be
\eqn\susal{
\{ Q^{AA'}, Q^{BB'} \} =\epsilon^{AB} \epsilon^{A'B'} P^+~,
}
where $P^+= -i \partial/ \partial \sigma^-$.

Let us now consider the different supermultiplets that appear in the
model. We will use $\eta_+^{AA'}$ to parameterize the $(0,4)$
supersymmetry transformations.  The
$b^{AY}$ and $\psi_-^{A'Y}$ form a standard multiplet which transforms as
\eqn\stdmult{
\delta b^{AY} = i \epsilon_{A'B'} \eta_+^{AA'} \psi_-^{B'Y}, \qquad
\delta \psi_-^{A'Y} = \epsilon_{AB} \eta_+^{AA'} D_= b^{BY}~.
}
Here we define the covariant derivative by
$D_\mu = \partial_\mu - g A_\mu^\alpha T_R^\alpha$, where $T_R^\alpha$ is the 
gauge generator for the representation $R$.
The $b^{A' \tilde A'}$ and $\psi_-^{A \tilde A'}$ form a twisted
multiplet
which transforms as
\eqn\twistmult{
\delta b^{A'\tilde A'} = i \epsilon_{AB} \eta_+^{AA'} \psi_-^{B\tilde
A'},
\qquad
\delta \psi_-^{A\tilde A'} = \epsilon_{A'B'} \eta_+^{AA'} D_=
b^{B'\tilde A'}~.
}
The fields $\phi^{A' m}$ and $\chi_-^{Am}$ also form a twisted
multiplet and transform as \twistmult.

The gauge multiplet consists of $A_\mu$, $\psi_+^{A'A}$ and
$\psi_+^{\tilde A' Y}$. We will write $A_\mu$ in light-cone
coordinates in terms of components $A_\pp$ and $A_=$.
The terms in the Lagrangian and the supersymmetry transformations
involving these fields will be determined by the
Noether procedure. Namely, we begin with an action with just a $U(1)$
gauge symmetry and an $O(n)$ global symmetry and then add
terms
to the Lagrangian and supersymmetry transformations order-by-order in
the gauge coupling constant $g$ to obtain
a theory with $(0,4)$ supersymmetry and $O(n)$ gauge invariance.
One obtains the following result for the supersymmetry transformations
\eqn\gaugemult{
\eqalign{
\delta A_\pp &= -i \eps_{AB} \eps_{A'B'} \eta_+^{AA'} \psi_+^{BB'} \cr
\delta A_= &=0 \cr
\delta \psi_+^{AA'} &= F_{\pp =} \eta_+^{AA'} + g D^{AA'}_{BB'}
\eta_+^{BB'} \cr
\delta \psi_+^{\tilde A' Y}& = g D^{\tilde A' Y}_{BB'} \eta_+^{BB'}
{}~,\cr}
}
where we have defined
\eqn\dterms{
\eqalign{
D^{AA'}_{BB'} &= b^{AY} T_S b_{BY} \delta^{A'}_{B'} +
b^{A' \tilde A'} T_S b_{B' \tilde A'}  \delta^A_B +
\phi^{A' m} T_F \phi_{B'm}  \delta^A_B \cr
D^{\tilde A' Y}_{BB'} &= 2  b_A^Y T_S b^{\tilde A'}_{B'} ~,\cr}
}
and $S(F)$ refers to the symmetric (fundamental) rep.
The field strength is defined as  $F_{\pp =} = \partial_\pp A_= -
\partial_= A_\pp -g [A_=, A_\pp]$.

The purely left-moving fermionic multiplets are
$\chi_+^{Ym}$ and $\lambda_+^M$,
and we will frequently denote these as
$\lambda^a_+ = (\chi_+^{Ym},\lambda_+^M)$.  The supersymmetry
transformations
of these fields are determined in terms of the functions
$C^a_{AA'}$ of the bosonic fields as
\eqn\leftsusy{
\delta \lambda^a_+ = \eta_+^{AA'} C^a_{AA'}~.
}
The functions $C^a_{AA'}$ have global $O(n)$ symmetry. Since the terms in the
Lagrangian dependent on the $C^a_{AA'}$ do not contain derivatives,
they do not alter the supersymmetry transformations \gaugemult.
The functions $C^a_{AA'}$ may be determined by demanding $(0,4)$
supersymmetry \refs{\witadhm}. We require that up to gauge
transformations
\eqn\sustrans{
(\delta_{\eta'} \delta_\eta- \delta_\eta \delta_{\eta'} ) \lambda_+^a=
-i \epsilon_{A'B'} \epsilon_{AB} \eta_+^{AA'} \eta_+^{BB'} D_=
\lambda_+^a
= -i \epsilon_{A'B'} \epsilon_{AB} \eta_+^{AA'} \eta_+^{BB'}
G_\theta^a \rho^\theta~,
}
where $\rho^\theta$ includes all the right-movers, and $G^a_\theta$ is
a function of the bosonic scalars. The covariant derivative
$D_\mu =\partial_\mu - gA_\mu$ when acting on fields in the
fundamental representation. Equation \sustrans\ implies the condition
\eqn\susycond{
0= {\partial C^a_{AA'} \over \partial b^{BY} } +
{\partial C^a_{BA'} \over \partial b^{AY} } =
 {\partial C^a_{AA'} \over \partial \phi^{B'm} } +
{\partial C^a_{AB'} \over \partial \phi^{A'm} } =
 {\partial C^a_{AA'} \over \partial b^{B'\tilde B'} }
+{\partial C^a_{AB'} \over \partial b^{A'\tilde B'} } ~.
}
The part of the Lagrangian containing $\lambda_+$ is then determined
to be
\eqn\lamlag{
\int d^2 \sigma \biggl( {i\over 2} \lambda^a_+ D_= \lambda^a_+ -
{i\over 2} \lambda_+^a G_{a\theta} \rho_-^\theta \biggr)~,
}
with
\eqn\gfuncis{
G^a_\theta \rho^\theta_- = {1\over 2}\biggl(
\epsilon^{BD} { {\partial C^a_{B B'}} \over {\partial b^{DY}}}
\psi_-^{B'Y}
+ \epsilon^{B'D'} {{\partial C^a_{B B'}} \over {\partial \phi^{D'Y'}}}
\chi_-^{B Y'}  +
\epsilon^{B'D'} {{\partial C^a_{B B'}} \over {\partial b^{D'\tilde D'}}}
\psi_-^{B \tilde D'}
\biggr)~.
}

To obtain a Lagrangian with the full $(0,4)$ supersymmetry it is
useful to compare to the case with $(0,1)$ supersymmetry where a
superfield formalism is available. We have
\eqn\oonesusy{
\delta \lambda^a_+ = \eta F^a~,
}
where $F$ is an auxiliary field. The potential energy of the system is
$\sum_a (F^a)^2/2$. On shell $F^a$ should be identified with $C^a$, say
as $F^a = c^{AA'} C^a_{AA'}$, so
to obtain a theory with $(0,4)$ supersymmetry we must require that $V$
be independent of $c^{AA'}$. Note $c$ should be normalized so that
$c^2 =1$. This leads to the condition
\eqn\susycondf{
0= \sum_a {\rm Tr}(C^a_{AA'} C^a_{BB'} + C^a_{BA'} C^a_{AB'} )~.
}

Solving conditions \susycondf\ and \susycond\ yields
\eqn\csol{
\eqalign{
C^M_{AA'} & = g h^M_{Am} \phi_{A'm}, \qquad
C^{Ym}_{AA'} = g \phi_{A'n} (X^{AY}_{mn} - b^{AY} \delta_{mn} ) ~.\cr}
}
Here $h$ and $X$ are fields in the D-fivebrane worldvolume theory, which
appear as background fields from the point of view of the
D-strings. These
fields must satisfy the D-flatness conditions of the fivebrane
worldvolume theory,
\eqn\dflat{
h_m^{M(A} h_{Mn}^{B)} + \eps^{pq} \eps_{YZ} X_{mp}^{(AY} X_{nq}^{B)Z}
=0~.
}

The final Lagrangian is given by
\eqn\lagrangian{
\eqalign{
L &= {1\over 2} \int d^2\sigma  {\rm Tr} \biggl( D_\pp b^{AY} D_=
b_{AY}+D_\pp b^{A'\tilde A'} D_=b_{A' \tilde A'} + D_\pp
\phi^{A'm} D_= \phi_{A'm} \cr &-i \rho_-^\theta D_\pp \rho_{-\theta} +
i \lambda_+^a D_=
\lambda_{+a} - F_{\pp =}^2  + i \psi_+ D_= \psi_+ \biggr) \cr
&
- {i\over 4} \int d^2 \sigma {\Tr }~\lambda_{+a}\biggl(
\epsilon^{BD} { {\partial C^a_{B B'}} \over {\partial b^{DY}}}
\psi_-^{B'Y}
+ \epsilon^{B'D'} {{\partial C^a_{B B'}} \over {\partial \phi^{D'm}}}
\chi_-^{B m}
\biggr) \cr &- {1\over 8} \int  d^2 \sigma {\Tr }\biggl(
\epsilon^{AB} \epsilon^{A'B'} C_{aA A'} C^a_{B B'} \biggr) \cr
& -ig \int d^2 \sigma \Tr \biggl(  b_{AY}\psi_+^{AA'} \psi^{Y}_{-A'} +
b_{A'\tilde A'}\psi_+^{AA'}  \psi^{A}_{-\tilde A'}+
\phi_{A'm}\psi_+^{AA'}  \chi^{m}_{-A} \cr
&+b_{AY}\psi_+^{\tilde A' Y}  \psi^{A}_{-\tilde A'} +
b_{A'\tilde A'}\psi_+^{\tilde A' Y}  \psi^{A'}_{-Y} \biggr)
\cr & - {g^2\over 4} \int d^2 \sigma \Tr \eps^{BC} \eps^{B'C'} \biggl(\eps_{AD}
\eps_{A'D'} D_{BB'}^{AA'} D_{CC'}^{DD'} + \eps_{\tilde A' \tilde B'}
\eps_{YZ} D_{BB'}^{\tilde A' Y} D_{CC'}^{\tilde B' Z} \biggr)~.\cr
}
}
When the number of D-fivebranes $k$ vanishes, and one introduces an
$SO(32)$ Wilson line, which makes half of the $\lambda_+^M$ periodic
and the other half anti-periodic, this reduces to the
two-dimensional
$(0,8)$
system considered in \lowetwo. When the number of D-strings is one,
this
reduces to the linear sigma model considered in \refs{\witadhm, \dougtwo}.

As discussed in the introduction, the small instanton limit
corresponds
to taking the Type IA coupling constant to infinity. This translates
into taking the coupling $g$ of \lagrangian\ to infinity, which
in turn corresponds to taking the infrared limit of the theory.
To describe the uncompactified case the size of the $\sigma$
direction should be sent to zero.
When combined
with taking the large $n$ limit and setting $X$ and $h$ to zero,
this gives a matrix description of
small instantons in $E_8\times E_8$ heterotic string theory.

For
finite
$n$ we will argue below this gives a description of $E_8\times E_8$
small
instantons, or equivalently gauge fivebranes of heterotic string
theory, in which the longitudinal direction parallel to the fivebranes has been
compactified. Taking $n$ finite then corresponds to considering the
discretized light-cone quantization of this theory. We will elaborate
on the details of this further in the following sections.

Note that
although the $X$ and $h$ couplings vanish for the case of small
instantons, retaining these couplings will clarify the infrared behavior
of the theory, as we will see later. The $X$ fields describe the
positions
of the D-fivebranes in the four transverse dimensions. The additional
degrees of freedom that describe the positions of the M-theory
fivebranes in the eleventh dimension arise when one allows for
nontrivial boundary conditions for the $\phi^{A'm}$ fields as one goes
around $\sigma$. These are the remnants of the Wilson line degrees of
freedom of the $Sp(2k)$ gauge group on the worldvolume of the
fivebranes.

If the radius of the $\sigma$ direction is held fixed in the infrared
limit, the theory \lagrangian\ gives a matrix formulation of a new noncritical
six-dimensional string theory with $E_8\times E_8$ global symmetry
and $(1,0)$ supersymmetry,
analogous to the $(2,0)$ six-dimensional noncritical string theories
recently considered in \seibergthree.

\newsec{Matrix Theory Approach}

Let us see how the gauged linear sigma model found in the previous section
may be derived from the Matrix theory approach.
Our starting point is the heterotic Matrix model
\refs{\daniel,\kacsil,\lowe}. Compactifying on a circle one obtains
a two-dimensional $O(n)$ gauge theory as described in \refs{\banks,\lowetwo}.
Compactifying further on a $T^4$ one obtains a six-dimensional
non-critical
string
theory discussed
in \refs{\seibergthree,\dvvtwo} compactified on $K3\times S^1$. The
moduli space of vacua in this theory will be $SO(20,4,\IZ) \backslash SO(20,4)
/ (SO(20)\times SO(4))$, and the duality group (including mirror
symmetry) is identified with $SO(20,4,\IZ)$.
We work at a point where the $K3$ is $T^4/\IZ_2$ where the $T^4$
has sizes $\Sigma_i$ ($i=1,\cdots,4$) and the $S^1$ has size
$\Sigma_5$, with all angles right angles and the three form set to
zero.
The parameters of the
compactified $(2,0)$ theory are identified with parameters of the
M-theory on a dual $T^4/\IZ_2\times S^1$
(with sizes $L_i$ with $i=1,\cdots,5$) via
\eqn\mdual{
\eqalign{
\Sigma_i &= {l_p^3 \over {R L_i}} , \quad (i=1,\cdots,4)\cr
\Sigma_5 &= {l_p^6 \over {R L_1 L_2 L_3 L_4}} \cr
M_s^2 &= {{R^2 L_1 L_2 L_3 L_4 L_5} \over l_p^9} ~,\cr}
}
where $R$ is the length of the longitudinal direction, $l_p$ is the
eleven
dimensional Planck length, and $M_s$ is string scale of the
noncritical string theory.

When $M_s \to \infty$ this reduces to the $(2,0)$ field theory (see
\seibergsusy\ for a review) compactified on $K3 \times S^1$, as studied
in \berkooz. This should correspond to the heterotic matrix model
compactified on $T^3$.

In general, this noncritical string theory only looks like a
supersymmetric field theory
in special corners of the moduli space. Actually, we will chiefly
be interested in the limit when the $T^4$ is very large
(i.e. $L_1,\cdots,L_4$ large) with the eleven dimensional Planck
length fixed which
precisely corresponds to one of these corners.
At the point with $SU(2)^{16}$
enhanced gauge symmetry this theory flows in the infrared to
a $U(n)$ Yang-Mills theory
on a dual orbifold space $S^1 \times T^4/\IZ_2$. To relate this to the
heterotic Matrix model we need to consider a limit in which one
direction of the $T^4$ is much smaller than the other length scales.
After Kaluza-Klein reduction
one obtains a Yang-Mills theory on $S^1\times T^3/\IZ_2$, with certain
boundary conditions on the fields:
\eqn\fident{
\eqalign{
A_{0,1} (\sigma, \sigma^i) &= -A^{\dag}_{0,1} (\sigma, -\sigma^i) \cr
A_a(\sigma, \sigma^i) &= A^{\dag}_a (\sigma, -\sigma^i) \cr
X_k(\sigma, \sigma^i) &= X_k^{\dag}(\sigma, -\sigma^i) ~,\cr }
}
together with extra 1+1-dimensional fermionic degrees of freedom that live at
the fixed points \refs{\berkooz,\horavatwo}.
We wish to introduce an instanton background into this theory and
then integrate out the heavy modes to obtain the effective action.
To do this we decompose $U(n)=U(n_0)\times U(n_1)$ and embed the
instanton in the $U(n_0)$ factor. The scalars decompose as
\eqn\scalard{
X_k = \biggl( \matrix{ Z_k & Y_k \cr
                       Y^{\dag}_k & x_k \cr} \biggr)~,
}
and a similar equation holds for the fermions. Here $Z_k$ is a $n_0
\times n_0$ matrix, $x_k$ is $n_1\times n_1$, etc.
The leading contribution to the
action comes from the zero modes of the $Y$ field (and its fermionic
partner) in the background of the instanton. For a $k$ instanton
configuration there will be $k n_1$ such zero modes, which will
transform in a bifundamental representation $(k,n_1)$.
In
the limit that the size of the dual torus $T^3$ shrinks to zero size we
obtain
a description of a small instanton in flat space, and the matrix
description
reduces to a two-dimensional field theory. For $k=0$ one finds the
usual fields of the heterotic Matrix theory \lowetwo. For $k\neq 0$,
when one takes into
account the additional projection \fident\ one finds precisely the
additional
worldsheet fields $\phi^{A'm}$ together with the fermionic fields
$\chi_+^{Ym}$ and $\chi_-^{Am}$ of the previous section. These pick up
an additional sign under the $\IZ_2$ projection with respect to the
$b^{AY}$ scalars, so the new zero modes have an $Sp(2k)$ global
symmetry, rather than $O(2k)$. Gauge
symmetry and $(0,4)$ supersymmetry then fixes the effective action to
be of the form \lagrangian.

\subsec{Discrete Light-Cone Quantization}

Now let us make a few comments on the interpretation of this
Lagrangian for finite $n$. The comments we make here are also relevant
for the theory without the instanton background.
In Type \II\ case it has been conjectured the finite $n$ Matrix
theory describes a DLCQ sector of M theory with the longitudinal
direction $x^-$ compactified on a circle of radius $R$.
The same argument can be made in this case. A new feature now
is the presence of Wilson line degrees of freedom corresponding to the
boundary conditions on the gauge fermions. In general, these Wilson
lines are unconstrained. If however we want to de-compactify, by taking
a large $n$ limit, the Wilson line must be carefully chosen if we are to
obtain a theory with a conventional local spacetime interpretation
and $9+1$-dimensional Lorentz invariance. Apparently,
the only consistent way to do this is to take the Wilson line as in
\lowetwo\ which gives half of the $\lambda_+^{M}$ fields periodic
boundary conditions, and the other half antiperiodic boundary
conditions as one goes around the $S^1$. The absence of manifest
$E_8\times E_8$ invariance is then simply due to the presence of this
Wilson line at finite radius. Similar comments have also recently appeared
in
\motl.

\newsec{Spectrum of States}

As discussed above, the theory \lagrangian\ for finite $n$
can be used to describe small instantons compactified on a
longitudinal circle in
the presence of a Wilson line.
The large $n$ limit of the theory is conjectured to describe the
full uncompactified six-dimensional theory associated with small
$E_8\times E_8$ instantons. In the following we will consider the case
of small instantons in a single $E_8$ factor.
This theory may alternatively be studied
by considering F-theory compactified on an elliptic Calabi-Yau
threefold
with a collapsing del Pezzo surface. Compactifying further on a circle
yields M-theory on a similar Calabi-Yau, giving a description of the
4+1-dimensional theory.
The particle-like BPS states that appear from this point of view were
studied in
\refs{\ganor,\klemm} by counting cycles in Calabi-Yau manifolds.

These results can
be summarized in terms of a non-critical string theory with
$E_8$ quantum numbers.
For the case of single winding number of this
non-critical string, they find the result that the degeneracy $d(n_E)$ of
states with momentum $n_E$ is given by
\eqn\eeightd{
q^{-{1\over 2}} \sum_{n_E=0}^\infty d(n_E) q^{n_E} = {\theta_{E_8}(q) \over
{\eta(q)^{12}}}~.
}
The $SO(4)_I=SU(2)\times SU(2)$
spacetime quantum numbers of these
states may also be deduced.
For $n_E=0$ they find a
hypermultiplet singlet of $E_8$. The hypermultiplet has spacetime
quantum numbers $4(0,0)\oplus(\half,0)\oplus (0,\half)$.
The states with $n_E=1$ and their CPT conjugates form
a hypermultiplet in
the ${\bf 248}$ together with a singlet in the
$4(\half, \half) \oplus (1,\half)
\oplus (\half,1) \oplus (0,\half) \oplus (\half,0)$. The states were
found by considering the limit in which the del Pezzo surface
collapsed to zero size. Precisely at the critical point it is possible
for the states with $n_E>1$ to decay, for example to states with
winding number $\pm 1$ and $n_E=0, \pm 1$.

To compare directly with the Matrix theory results for
finite $n$ it is
necessary to introduce a Wilson line around the compact direction
which breaks the $E_8$ symmetry to $SO(16)$.
This splits the ${\bf 248}$ of $E_8$ into the ${\bf 120}$ of
$SO(16)$ and the ${\bf 128}$.
The masses of these states change in the usual way
\eqn\wilmass{
M = | { {n_E +A_E P_E} \over 2R}|~,
}
where $A_E$ is the Wilson line parameter in $E_8$ variables, and $P_E$
is a point on the $E_8$ weight lattice.

In general, the relations between the $E_8$ quantities  and the
$SO(32)$ quantities that appear in the Matrix theory are
\eqn\dict{
n = 2n_E + 2A_E P_E +  A_E^2 m_E, \quad 2m + 2A P + A^2 n = m_E~,
}
where $(n,m)$ are the winding number and momentum of a fundamental $SO(32)$
string, $A$ is the Wilson line
$(\half^8,0^8)$ in the Cartan subalgebra of $SO(32)$,
and $P$ is
a point on the lattice $Spin(32)/\IZ_2$. Here $m_E$ is the winding
number of a fundamental $E_8 \times E_8$ string, which is to be
identified with the worldsheet momentum of \lagrangian. At present we
consider the limit that the radius of the $\sigma$ direction vanishes,
so we set $m_E=0$.

The formula \dict\ implies the singlet with $n_E=0$  lies at
$n=0$. The ${\bf 120}$ of $SO(16)$ at $n_E=1$ will lie at $n=2$
with $2(0,0)\oplus (\half,0)$ $SO(4)_I$ quantum numbers. The
singlet at $n_E=1$
will lie at $n=2$ with $2(\half, \half) \oplus (1,\half) \oplus(0,\half)$
quantum numbers. The ${\bf 128}$ at $n_E=1$ will lie at $n=1$ with
$2(0,0)\oplus (\half,0)$ $SO(4)_I$ quantum numbers.
The higher representations of $E_8$ found in \eeightd\ will begin
to appear at $n=2$.

The Kaluza-Klein compactification of the six-dimensional field
theoretic states gives rise to additional states in five dimensions,
which are independent of the wrapped tensionless string
states.\foot{In \klemm\ it was noted this spectrum
depends on the details of the Calabi-Yau. The Matrix theory
with $k=1$ should correspond to the critical point in the transition
between the Higgs branch with 29 massless hypermultiplets and the
Coulomb branch with a single tensor multiplet.}
The tensor multiplet reduces to a tower of states with
$(1,0)\oplus (0,0) \oplus 2(\half,0)$ $SO(4)_I$ quantum numbers for
all $n_E$, which translates to all even $n$. The 29 massless
hypermultiplets of the six-dimensional theory transform as the
${\bf \half 56 +1}$ of the unbroken $E_7$ of the Higgs branch. When we
compactify and turn on Wilson lines, the $E_8$ is broken to
$SO(16)$. The $SO(16)$ is further broken to $SO(12)\times SU(2)$ by
the finite size instanton. Likewise the $E_7$ symmetry of six
dimensions will be broken to $SO(12)\times SU(2)$ by the Wilson lines,
under which ${\bf 56 \to (12,2) \oplus (32,1)}$. At the point
where the instanton shrinks to zero size, these representations will
be enlarged to representations of $SO(16)$. The ${\bf 32}$ will be
enlarged to a spinor ${\bf 128}$ which will generate a tower of
Kaluza-Klein states with $n$ odd, and $2(0,0)\oplus (\half,0)$
$SO(4)_I$ quantum numbers. Finally the singlet and the ${\bf (12,2)}$ states
will arise from the ${\bf 120}$ of $SO(16)$, which will generate a
tower of states with $n$ even, and $2(0,0)\oplus (\half,0)$ quantum numbers.

\subsec{$n=1$, $k=1$}

We first examine the instanton number one case, and set $n$ to one.
To deduce the BPS states from the Matrix approach,
we consider the conformal field theory
that \lagrangian\ flows to in the infrared.
The potential energy appearing in \lagrangian\ takes the form
\eqn\poten{
V = \sum |C|^2 \sim \phi^2 (X^2+b^2)~.
}
For generic ADHM data (i.e. D-fivebrane worldvolume fields $X$ and
$h$)
there is
one branch of the moduli space along which $\phi$ is massive.
In the infrared the Lagrangian will flow to one of
the wormhole
conformal field theories studied in \refs{\calharstr, \strom}.
These theories have $(0,4)$ worldsheet supersymmetry.
Naively
one might think these would be symmetric under a
$SO(4)=SU(2)\times SU(2)$ R-symmetry
group, but in fact the $\CN=4$ superconformal algebra is only symmetric
under a $SU(2)$ subgroup. On the branch where $\phi$ is massive the
$SU(2)$ that appears in the superconformal algebra is $SU(2)_{A'}$,
and
$SU(2)_A$ is spontaneously broken.

When $X=0$ the gauged linear sigma model describes an instanton of zero size.
At this point, a second branch of the moduli space appears, along
which
$b$ becomes massive and $\phi$ takes on a nonzero expectation value.
In this case, the linear sigma model flows to a different
superconformal
field theory in the infrared where now $SU(2)_A$ appears in the
superconformal algebra. Once again one may argue the conformal field
theory that appears is one of the wormhole conformal field theories
\refs{\calharstr, \strom}.

As discussed in \witcom\ a potential contradiction arises in this
picture. Naively the two branches meet at $b=\phi=0$ which would be
inconsistent with the origin of the R symmetry of the
superconformal algebra as discussed above. This problem is resolved if in the
conformal field
theory limit the distance to the $b=\phi=0$ point moves off to
infinity allowing the two branches to remain separate.

Let us consider further the conformal field theory that appears in
the infrared. One way to study this theory is simply to solve the
low-energy spacetime equations of motion order by order in $\a'$. To
leading order this yields
\eqn\metric{
ds^2 = (db)^2 (e^{2\phi_0} + 8 \alpha' { (b^2 + 2X^2) \over
(b^2+X^2)^2} + O({\a'}^2) )~,
}
where $\phi_0$ is the value of the dilaton at infinity and $(db)^2$ is the
usual
flat space metric. As $X\to 0$ the spacetime develops a long tube,
consistent with $b=\phi=0$ being at infinite distance in the conformal
field theory. It is possible to show this solution may be corrected
order by order in $\a'$ to yield an exact conformal field theory
\chstwo. Unfortunately little is known about this exact conformal
field theory. The next to leading order solution for the spacetime fields
obtained from the ADHM sigma model has been obtained in \lambert,
where it is shown the solution differs from the instanton solution
of \calharstr. Nevertheless the corrections remain consistent with the
$b=\phi=0$
point being at infinite distance.

The construction of the states in the exact conformal field theory
is a difficult problem. One approach would be to do a Witten index
calculation, however because the ground states will appear as bound states at
threshold, this index calculation is rather subtle. Instead we will
show the states expected arise in two different limits where we may
use algebraic CFT techniques to analyze the theory.

The first limit we consider is the CFT describing the region far from
the wormhole, which we may think of as the Coulomb branch of the
two-dimensional theory.
This approaches flat space, so the CFT is the usual free theory
heterotic theory with $(0,8)$ supersymmetry.
The spectrum of states
is the same as in \lowetwo\ so for $n=1$  one finds the ${\bf 128}$
of $SO(16)$.  This
lies in the ${\bf 8_V\oplus 8_S}$ of $SO(8)$ which decomposes into
$(\half,\half)\oplus 4(0,0)\oplus 2(\half,0)\oplus 2(0,\half)$ under
$SO(4)_I$. These are the usual states of the compactified
ten-dimensional heterotic string theory and are expected to decouple
from the intrinsically six-dimensional degrees of freedom associated
with the small instanton.

The second limit we consider, where we may still explicitly describe
the
CFT is the long tube region of the spacetime
that appears in the zero-size limit $X\to 0$. This can be described by a
tractable conformal field theory which is a  supersymmetric
$SU(2)$ WZW model together
with a free field with a linear dilaton \calharstr, and the usual
free left-moving gauge fermions $\lambda_+^M$. The level $k_w$ of
the WZW model is identified with the charge of the solution. We take
$k_w=1$ in the case at hand. With an appropriately defined energy
momentum tensor, the central charge of this theory is $6$ \jack.
The mass-shell condition is
\eqn\massshell{
m^2= -1 +\half P_L^2 + \half P_R^2 +{c_R+c_{\bar R} \over 2+k_w} + N+\bar N~,
}
where $P_L$ and $P_R$ are the left and right-moving momenta,
$c_R$ is the second Casimir of the representation $R$,
and $N$ and $\bar
N$ are left and right-moving oscillator numbers. The
level-matching
condition becomes
\eqn\levelmatch{
\half P_R^2+{c_{\bar R} \over 2+k_w} + \bar N =
 \half P_L^2+{c_{ R} \over 2+k_w} +  N-1~.
}
The BPS condition requires the right-movers to be in their ground
states.
The states found from the Calabi-Yau point of view are obtained
by taking $R$ and $\bar R$ to be singlets, and the derivation of the
spectrum proceeds as in the $(0,8)$ case, with identical results.

Now let us consider the Higgs branch of the two-dimensional theory
where
$\phi$ has a non-zero vev. The CFT that appears on
this branch is similar to the one just considered, but now orbifolded by the 
$\IZ_2$ symmetry that acts on $\phi$ and $\chi$. The quantum
numbers of the ground states arise from the quantization of the
fermion zero modes $\lambda_+^M$ and $\chi_-^{Am}$. The $Sp(2)$
quantum numbers are associated with the tensionless string winding
number. One finds the ${\bf 128}$ of $SO(16)$ in the
$2(0,0)\oplus(\half,0)$ of $SO(4)_I$ and in the $2(0,0) \oplus (\half,0)$ of
$Sp(2)\times SU(2)_{A}$, which contains the states expected from the
Kaluza-Klein compactification of the six-dimensional hypermultiplets,
and the wrapped tensionless strings.

We see therefore that in either limit, the conformal field theories
contain all the states predicted from the Calabi-Yau approach.
Of course, we are really interested in normalizable states of the exact
conformal field theory which interpolates between these two limits.
It is expected the additional states we have found will turn out to be
non-normalizable when one constructs their wavefunctions globally.
It would be very interesting to do a Witten index calculation to
confirm this.

\subsec{$k=1$, $n>1$}

Now we consider the case where we have multiple D-strings and a single
D-fivebrane. Naively one might think the picture would be
similar to \lowetwo, namely that in the infrared the theory would flow
to a symmetrized product of the
conformal field theories discussed in the previous subsection.
The difference here is that now there exist marginal deformations of
the CFT consistent with the $SO(4)_I$ rotational invariance and gauge
symmetry. Following \withig\ where the fivebrane of Type \II\ was
considered, we may argue there is a unique marginal operator which
preserves the $SO(4)_E$ group of rotations transverse to the
fivebrane.
In \withig\ the coupling of this marginal operator was fixed by demanding
that a worldsheet theta angle vanish. Presumably a similar argument
will apply here. The undeformed orbifold conformal field theory leads
to a non-vanishing value for the worldsheet theta angle, so does not
describe the correct infrared fixed point for the gauge theory.

If one is only interested in the spectrum of
BPS states, one could argue these are
insensitive to the deformation, allowing computations to be
performed
at the orbifold point.
It is unlikely one can trust this argument in detail for the present
case. However let us
follow this logic for the case $n=2$ and apply a Born-Oppenheimer
approximation as in the previous subsection. For the singlet and
${\bf 120}$ representations of $SO(16)$ one does indeed find
results consistent with states expected from Kaluza-Klein
compactification of the six-dimensional field theory, together with
wrapped tensionless string states predicted by the Calabi-Yau approach.
However, there is no
sign
of the higher representations of $E_8$ predicted by \refs{\klemm ,
\ganor}. It is possible these arise as bound states at threshold once
the interactions induced by the marginal operator
are properly included, or alternatively they may only be stable on the
Coulomb branch of the six-dimensional theory.
Similar arguments can be made for the $n>2$ case.

\subsec{$k>1$}

For many D-fivebranes, and a single D-string
\lagrangian\ will flow in the infrared to a wormhole conformal field
theory \calharstr, with level number $k_w=k$. The construction of the
BPS states proceeds as above for the case $n=1$.
For multiple D-strings one will again
find a symmetric product of these conformal field theories, up to
marginal deformations involving twist fields. Further
consideration of this case is beyond the scope of the present work.

\subsec{Coulomb Branch}

The six-dimensional $E_8$ theory also has a Coulomb branch where
the massless states are a single tensor multiplet. To see this branch
from the Matrix point of view we must set $X=h=0$ and turn on the
Wilson line degrees of freedom discussed at the end of section two.
These are identified with the position of the fivebrane relative to
the end of the world in the eleventh direction. The Lagrangian flows
to that of a Type \IIB\ theory with a D-fivebrane interacting with
$n/2$ D-strings (here we must set $n$ even). The $O(n)$ gauge symmetry
is broken to $U(n/2)$. For $n=2$ this configuration is T-dual to a
D-particle plus D-fourbrane system in Type \IIA. In \sethi\ this was
shown to yield a single bound state with $\CN=4$ vector multiplet
quantum numbers from the four-dimensional point of view. The $\CN=4$
vector multiplet decomposes into a $\CN=2$ vector multiplet and a
hypermultiplet. The $\CN=2$ vector is precisely what is expected by
compactifying the six-dimensional tensor multiplet. The hypermultiplet
represents the center of mass degree of freedom of the fivebrane which
decouples from the six-dimensional theory.

For $n>2$, the theory will flow to a $(4,4)$ conformal field theory of
the type considered in \withig. This will look like a symmetric
orbifold of the $n=2$ CFT's with a marginal deformation turned on.
If one is prepared to assume the BPS spectrum is
unchanged by the marginal deformation one can argue the requisite
bound states arise from the $\IZ_{n/2}$ twisted sector of the
undeformed orbifold.

\bigskip

\centerline{\bf Acknowledgments}

This work was supported in part by DOE grants DE-FG03-92-ER40701,
and DE-FG0291ER40688-Task A. I
wish to thank the Aspen Center for Physics for hospitality during the
completion of this work.

\listrefs
\end